\documentclass[aps,twocolumn,groupedaddress,showpacs]{revtex4}
\usepackage{graphicx}
\begin{document}
\bibliographystyle{apsrev}


\title{Manifestations of Broken Symmetry: The Surface Phases of Ca$_{2-x}$Sr$_{x}$RuO$_4$}


\author{R. G. Moore$^1$}
\author{V. B. Nascimento$^1$}
\author{Jiandi Zhang$^{2,3}$}
\author{J. Rundgren$^{4}$}
\author{R. Jin$^3$}
\author{D. Mandrus$^{3}$}
\author{E. W. Plummer$^{1,3,5}$}

\affiliation{$^1$Department of Physics and Astronomy, University of
Tennessee, Knoxville, TN 37996, USA}

\affiliation{$^2$Department of Physics, Florida International
University, Miami, FL 33199, USA}

\affiliation{$^3$Materials Science and Technology Division, Oak
Ridge National Laboratory, Oak Ridge, TN 37831, USA}

\affiliation{$^4$Department of Theoretical Physics, Alba Nova
Research Center, Royal Institute of Technology (KTH), SE-106 91
Stockholm, Sweden}

\affiliation{$^5$Center for Nanophase Materials Sciences, Oak Ridge
National Laboratory, Oak Ridge, TN 37831, USA}



\begin{abstract}
The surface structural phases of Ca$_{2-x}$Sr$_{x}$RuO$_{4}$ are investigated using quantitative Low Energy
Electron Diffraction. The broken symmetry at the surface enhances the structural instability against the RuO$_{6}$
rotational distortion while diminishing the instability against the RuO$_{6}$ tilt distortion occurring within the
bulk crystal. As a result, suppressed structural and electronic surface phase transition temperatures are
observed, including the appearance of an inherent Mott metal-to-insulator transition for $x = 0.1$ and possible
modifications of the surface quantum critical point near $x_{c} \sim  0.5$.
\end{abstract}

\pacs{68.35.Bs, 61.14.Hg, 74.70.Pq, 71.30.+h}
\maketitle

The discovery of superconductivity in Sr$_{2}$RuO$_{4}$ created a flurry of experimental and theoretical
activity\cite{maeno_nat94}. The structural similarity with La$_{2}$CuO$_{4}$, the parent compound of the
superconducting cuprates, combined with the nonconventional p-wave superconducting order parameter makes
Sr$_{2}$RuO$_{4}$ a focus of intense investigation \cite{maeno_pt01}. The substitution of Ca$^{2+}$ for Sr$^{2+}$
yields a phase diagram similar to the high-T$_{c}$ cuprates thus offering another opportunity to study the ground
state evolution from an antiferromagnetic Mott insulator to a superconductor \cite{nakatsuji_prl00,
nakatsuji_prl03, friedt_prb01}. One advantage of the Ca$_{2-x}$Sr$_{x}$RuO$_{4}$ [CSRO] compounds is isovalent
substitution between Ca$^{2+}$ and Sr$^{2+}$ which alters structural, electronic and magnetic properties by tuning
lattice distortions. Numerous theoretical and experimental works reveal the intricate coupling of the RuO$_{6}$
structural distortions with the electronic and magnetic degrees of freedom \cite{nakatsuji_prl00, nakatsuji_prl03,
friedt_prb01, braden_prb98, wang_prl04, fang_prb01, fang_prb04, dai_cm06, liebsch_prl07}. Another advantage is
that CSRO is a layered perovskite compound, thus its crystals are ameanable to cleaving. As such, the study of a
pristine [0 0 1] surface is possible through \textit{in situ} cleaving under ultra high vacuum conditions, thus
allowing an opportunity to investigate the intricate coupling between structure and other active degrees of
freedom in an environment of broken symmetry. In this work, the surface structural phases are determined by
quantitative analysis of Low Energy Electron Diffraction (LEED \textit{I-V}) spectra and compared to bulk studies
\cite{pendry74, vanhove86}.

\begin{figure}
\includegraphics[keepaspectratio=true, width = 3.2 in] {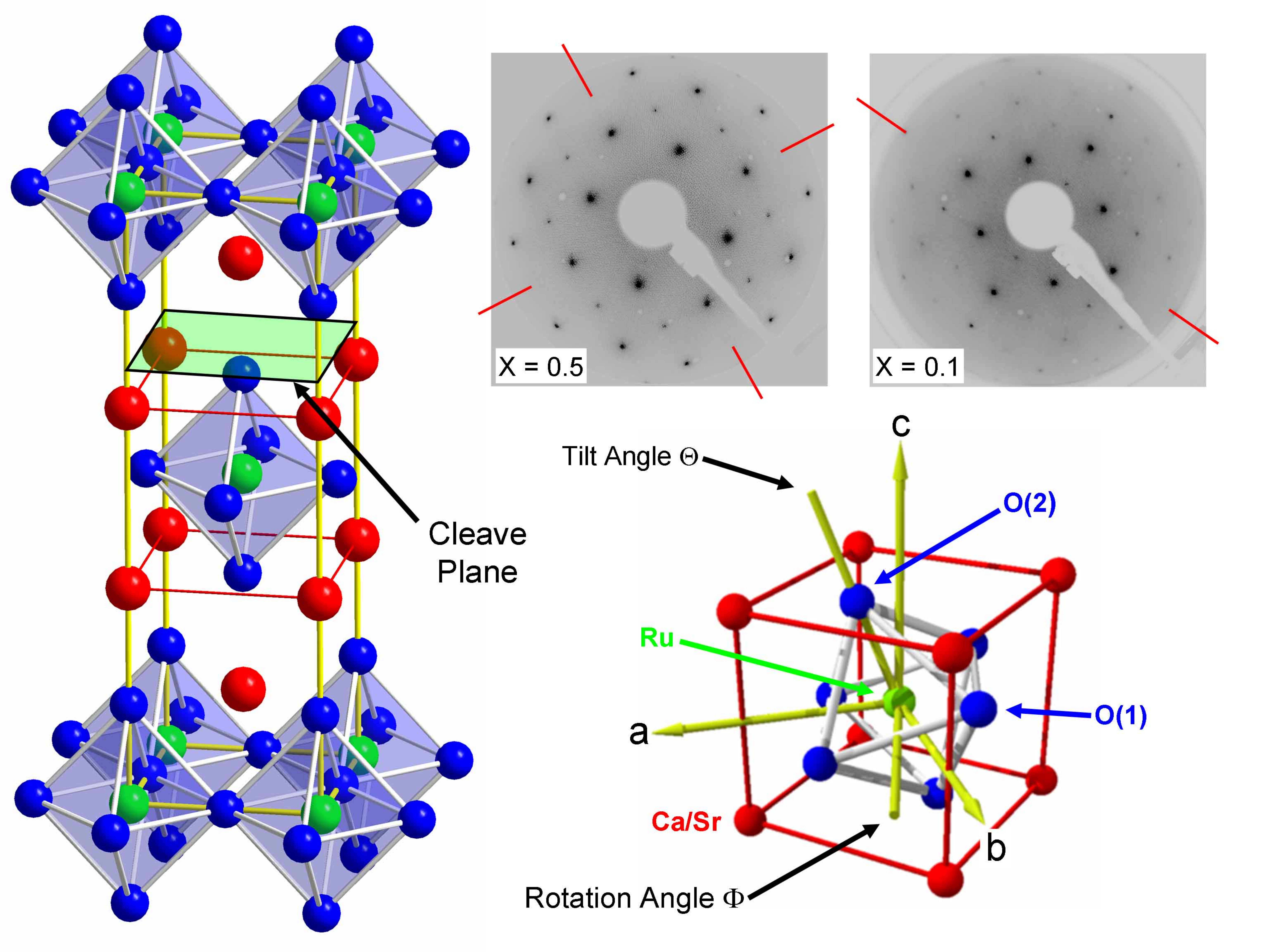}
\caption{Left: Bulk \textit{I4/mmm} structure of Sr$_{2}$RuO$_{4}$. Top Middle: LEED pattern for $x = 0.5$ showing
\textit{p4gm} plane group symmetry. Glide lines (on which fraction spots are extinct) are emphasized by red lines.
Top Right: LEED pattern for $x = 0.1$ showing \textit{pg} symmetry with only one glide line. Bottom Right:
Structural parameters used in describing bulk and surface geometries.} \label{PicturePhases}
\end{figure}

The bulk structural phases of the CSRO family have been previously determined by x-ray and neutron scattering
utilizing both powder and single crystal samples \cite{friedt_prb01, braden_prb98, braden_pc97, friedt_thesis}.
Starting from the highly symmetric \textit{I4/mmm} symmetry (no RuO$_{6}$ tilt or rotation) of Sr$_{2}$RuO$_{4}$
shown in Fig.~\ref{PicturePhases}, the smaller Ca$^{2+}$ cation shrinks the unit cell volume while the RuO$_{6}$
volume remains fairly constant. The shrinking cage surrounding the octahedron induces a chemical pressure rotating
the RuO$_{6}$ into an $I4_{1}$\textit{/acd} symmetry while maintaining a uniform octahedral shape and volume for
$0.5 \le  x \le  1.5$. When $x  < 0.5$, an octahedral tilt is induced entering into an orthorhombic \textit{Pbca}
symmetry. For $0.2 \le  x < 0.5$, a temperature ($T)$-dependent second order phase transition is observed with no
hysteresis \cite{friedt_prb01}. For Sr$_{2}$RuO$_{4}$, the system instability against the rotational distortion is
illustrated by a softening of the RuO$_{6}$ rotational $\Sigma _{3}$ phonon mode \cite{braden_prb98b}. A similar
structural instability for $x \sim 0.5$ is characterized by a softening of the RuO$_{6}$ tilting $\Sigma _{4}$
phonon mode \cite{moore_unpub}. Both the $I4_{1}$\textit{/acd} and \textit{Pbca} phases can be viewed as arising
from the freezing of the $\Sigma _{3}$ and $\Sigma _{4}$ modes respectively. For $x < 0.2$ the system is always
found in the \textit{Pbca} phase \cite{friedt_prb01}. Across the metal-to-insulator transition (MIT) for $x < 0.2$
a structural phase transition is encountered described by a flattening of the RuO$_{6}$ and larger lattice
distortions. While the bulk symmetry does not change across the MIT, the Ru-O(2) oxygen bond lengths decrease
$\sim 0.05$ {\AA} while the Ru-O(1) bond lengths increase $\sim 0.05$ {\AA}. In addition, the tilt of the
RuO$_{6}$ increases $\sim 5^{\circ}$ on average \cite{friedt_prb01}. The structural distortions yield a smaller
c/a-axis ratio in the insulating phase while the volume of the RuO$_{6}$ increases $\sim 3\%$.

High quality single crystals were grown using the optical floating zone technique. All crystals were well
characterized and concentrations verified by energy dispersive x-ray analysis. Crystals were cleaved and measured
{\it in situ} with a base pressure of $8 \times 10^{-11}$ torr revealing pristine [001] surfaces with large
micrometer terraces observed by STM. While it has been shown previously that the surface of Sr$_{2}$RuO$_{4}$
reconstructs to form a lower symmetry \cite{matzdorf_sci00}, for $x \le 1.5$ the crystals reveal a $p(1\times1)$
surface as shown in Fig. 1.  All available beams were collected at normal incidence and symmetrically averaged
yielding 16 nonequivalent beams for $x = 0.1$ and 11 nonequivalent beams for $0.2 \le x \le 2.0$. Total $I-V$
energy ranges varied slightly from surface to surface but all $I-V$ sets were $> 3000$ eV with the majority being
$> 3700$ eV. Theoretical $I-V$ curves generated for structural refinements employed a modified version of the
SATLEED program described elsewhere \cite{satleed, nascimento_prb07}. Due to the glide plane symmetry, simulated
annealing optimization algorithms were written taking advantage of bulk space group symmetry generators tailored
for each surface \cite{kirkpatrick_sci83}. In addition, the performance of the simulated annealing algorithms was
checked by manual grid searches for a few concentrations. Additional fit parameters were included to account for
possible asymmetric c-axis displacements that do not destroy the observed $p(1\times1)$ LEED pattern.  The Pendry
reliability factor $(R_{p})$ was used as a measure of agreement between theory and experiment \cite{pendry_jpc80}.
For all surfaces studied the refined surface structures yielded $0.19 \le R_{p} \le 0.28$ indicating excellent
agreement between theory and experiment.

All $0.2 \le x \le 2.0$ samples cleaved at room temperature (RT) exhibit a \textit{p4gm} plane group symmetry. The
glide lines presented in Fig.~\ref{PicturePhases} are due to the rotation of the RuO$_{6}$ about an axis parallel
to the c-axis. While the expected symmetry for a bulk terminated $I4_{1}$\textit{/acd} surface ($0.2 \le x \le
1.5$) is \textit{p2gg}, multiple terrace terminations generate the \textit{p4gm} symmetry \cite{nascimento_prb07}.
For $x < 0.2$, a \textit{pg} plane group symmetry is revealed, also shown in Fig.~\ref{PicturePhases}, reflecting
the symmetry of the bulk terminated \textit{Pbca} structure. The \textit{Pbca} symmetry is generated from a
rotation plus a tilt of the RuO$_{6}$. The tilt destroys one of the glide lines and thus only one is evident in
the LEED pattern shown in Fig.~\ref{PicturePhases}.

\begin{figure}
\includegraphics[keepaspectratio=true, width = 3.4 in] {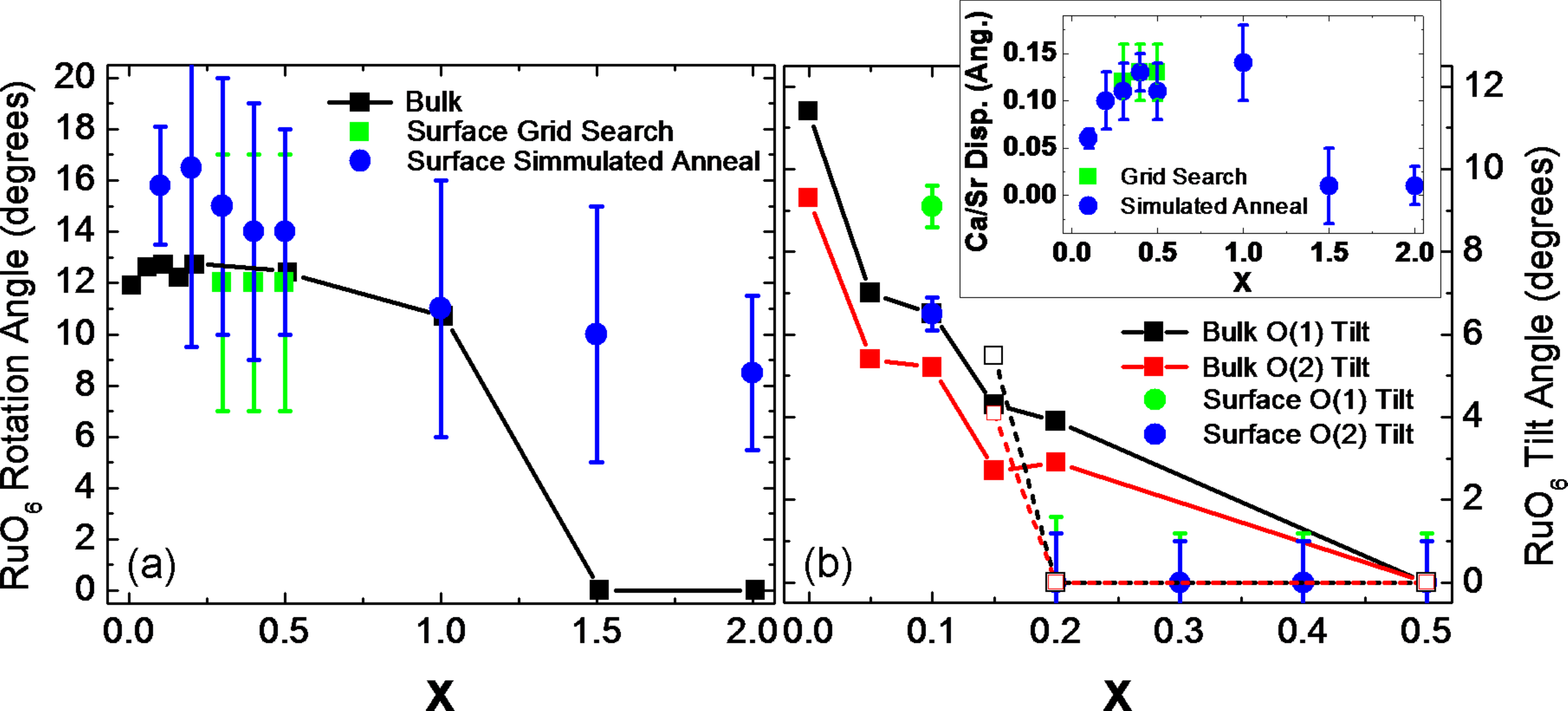}
\caption{(a) Bulk and Surface RuO$_{6}$ rotation angle versus concentration at $T = 300$ K. The surface equivalent
of a bulk terminated \textit{I4/mmm} symmetry never exists as a RuO$_{6 }$rotation exists for all $x$. (b) Bulk
and Surface RuO$_{6}$ tilt angles versus concentration. The solid lines (filled squares) represent bulk powder
data while the dashed lines (open squares) represent bulk single crystal data \cite{friedt_prb01, friedt_thesis}.
The critical tilt concentration is $x = 0.2$ for both bulk and surface single crystals. Tilt angles are slightly
enhance for $x = 0.1$. Inset shows the inward Ca/Sr displacement for different $x$. Vertical lines show transition
from \textit{p4gm} to \textit{pg} symmetry.} \label{RuO6Angles}
\end{figure}

The surface structures for the entire series at RT have been determined and the results are presented in
Fig.~\ref{RuO6Angles}. For $x > 1.5$, the bulk symmetry is \textit{I4/mmm,} however, \textit{no surface analog to
the I4/mmm symmetry (P4mm) is observed for any concentration.} The surface stabilizes the bulk instability against
the RuO$_{6}$ rotational distortion\cite{friedt_prb01}, freezing in the soft zone-boundary $\Sigma _{3 }$ phonon
mode creating a single \textit{p4gm} phase from $0.2 \le x \le 2.0$. The surface RuO$_{6}$ tilt angles at RT shown
in Fig.~\ref{RuO6Angles}b are more akin to bulk trends as no RuO$_{6}$ tilt is encountered for $x \ge 0.2$,
similar to bulk single crystal data \cite{friedt_prb01, friedt_thesis}. The tilts encountered for the $x = 0.1$
metallic phase are larger than those values encountered in the bulk metallic phase, but are smaller than the bulk
insulating phase. The largest surface relaxation observed on the CSRO surface involves the topmost Ca/Sr ions
where a significant inward motion is encountered for $x \le 1.0$ as shown in the Fig.~\ref{RuO6Angles}b inset. The
RT structure for $0.2 \le x \le 1.0$ shows a large $0.1$ {\AA} Ca/Sr inward motion but for $x = 0.1$, where a tilt
already exists, the inward motion is only $0.06$ {\AA}. A simple electrostatic argument would indicate that when
the surface is formed the topmost Ca/Sr-O(2) layer would be forced down \cite{moore_sci_unpub}, but the insert in
Fig.~\ref{RuO6Angles}b shows that it is not that simple. The surface buckling increases and is intimately tied to
the stability of the RuO$_{6}$tilt. While one might expect the creation of a surface to accentuate the system
instability against the tilt distortion, the observed trend discussed below indicates the RuO$_{6}$ tilt is
stabilized by the creation of a surface.

The RT LEED pattern for $0.2 \le x \le  2.0$ is shown in Fig.~\ref{PicturePhases}. The glide lines of the
\textit{p4gm} symmetry is evident by the extinguished ($\pm h,0$) and ($0,\pm h$) spots where $h$ is an odd
integer. To investigate the surface high temperature tetragonal-to-low temperature orthorhombic (HTT-LTO) phase
transition, crystals were cleaved at RT and subsequently cooled. As the \textit{Pbca} bulk phase boundary is
traversed the tilting RuO$_{6}$ octahedral destroys the glide line symmetry resulting in the appearance of the
($h,0$) beams. One would expect the low temperature LEED pattern to be similar to that of $x = 0.1$. However, such
is not the case as both the ($h,0$) and ($0,h$) beams are evident in the LTO LEED pattern revealing a {\it pm}
plane group symmetry.

\begin{figure}
\includegraphics[keepaspectratio=true, width = 2.4 in] {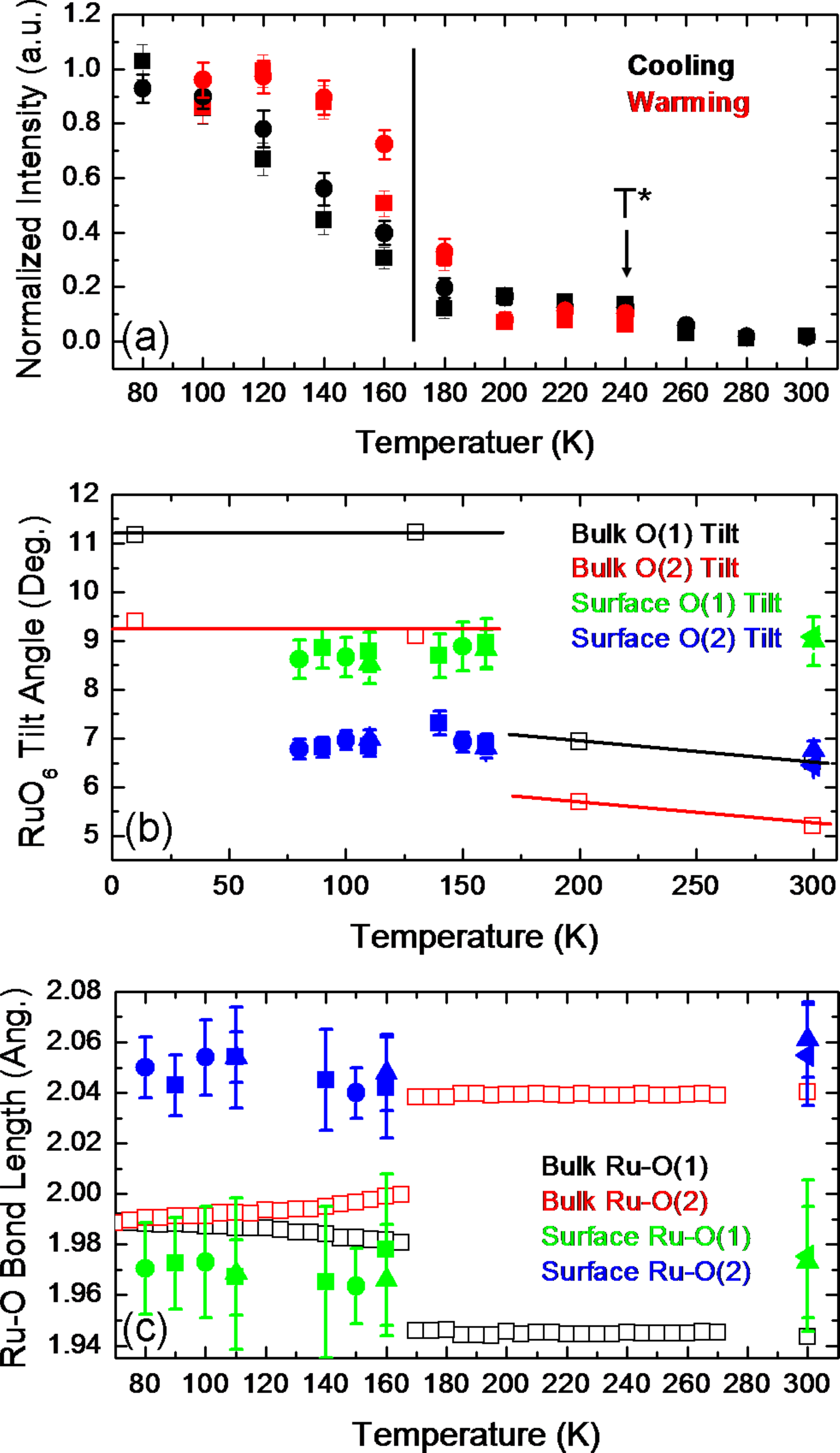}
\caption{(a) HTT-LTO phase transition order parameter for $x = 0.3$ showing first order phase transition character
with hysteresis. The solid squares are the integrated Beam ($0,3$) intensity at $176$ eV normalized to the Beam
($2,2$) intensity at $115$ eV. The solid circles are the normalized Beam ($3,0$) intensity. Vertical line shows
transition from \textit{p4gm} to \textit{pm} symmetry while arrow shows the onset of the tilt instability (T*).
(b) Surface RuO$_{6}$ tilt angles for $x = 0.1$ across bulk and surface MITs. The four closed symbols represent
four different crystal surfaces studied. The bulk data (open symbols) are shown for comparison with lines as
guides to the eye. The bulk data is from neutron powder experiments with a $T_{c} \sim 170$ K [5] while $T_{c}$ in
our bulk single crystals is $154$ K. The surface MIT $T_{c} = 130$ K\cite{moore_sci_unpub}. (c) Surface Ru-O bond
lengths for $x = 0.1$ across bulk and surface MITs. Neither the RuO$_{6}$ tilts nor the Ru-O bond lengths show
evidence (within experimental error) of a structural phase transition across the surface MIT.} \label{Ca19RuO6}
\end{figure}

\begin{figure}
\includegraphics[keepaspectratio=true, width = 3.2 in] {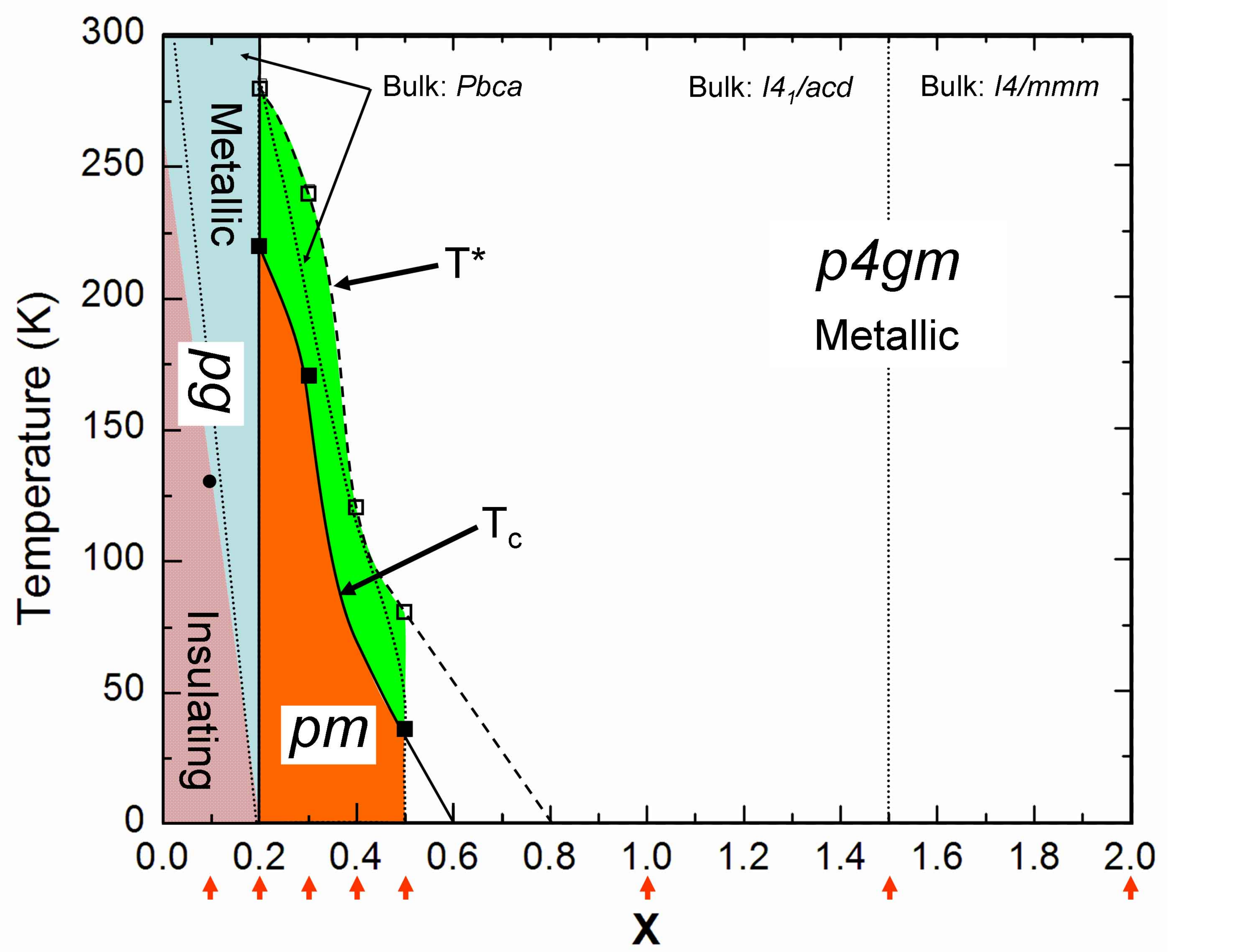}
\caption{Surface phase diagram for Ca$_{2-x}$Sr$_{x}$RuO$_{4}$. The dashed $T$* line is the temperature where a
tilt instability is revealed by weak diffuse reflections and the solid $T_{c}$ line is the \textit{p4gm} --
\textit{pm} structural phase boundary. The dotted lines represent bulk structural phase
transitions\cite{friedt_prb01,friedt_thesis,moore_unpub,jin_cm01}. There is no solid line between the metallic and
insulating phases for $x < 0.2$ as no structural phase transition exists across the surface MIT. Red arrows
indicate different concentrations investigated in this study. The unshaded regions below the solid and dashed
lines for $0.5 \le x \le  0.8$ are extrapolations based on observed surface trends.} \label{PhaseDiagram}
\end{figure}

Using integrated (0,3) and (3,0) beam intensity at $E_{i} = 176$ eV as an order parameter, the surface HTT-LTO
phase boundary is determined. As the system is cooled, broad diffuse (0,3) and (3,0) beams become evident for $0.2
< x \le  0.5$, indicated in Fig.~\ref{Ca19RuO6} and ~\ref{PhaseDiagram} by a temperature $T^{*}$. Such diffuse
beams are typical of short-range correlations similar to those observed in neutron data \cite{friedt_prb01,
friedt_thesis}. In contrast to neutron studies, the beam intensity is nearly constant for a considerable
temperature range indicating the system instability against the tilt distortion but never achieving the
\textit{Pbca} phase. As the phase boundary is traversed, the beam intensity dramatically increases and the beam
size shrinks as long range order is established. The behavior of both sets of beams is similar across the phase
boundary and beam intensity is the only difference as shown in Fig.~\ref{Ca19RuO6}. The normalized order parameter
intensity across the phase boundary for $x = 0.3$ is shown in Fig.~\ref{Ca19RuO6}a revealing $T_{c}\sim 170$ K,
some $20$ K below the bulk value \cite{jin_cm01}. While previous bulk studies demonstrate the lack of hysteresis
indicating a second order nature for the bulk phase transition \cite{friedt_prb01, moore_unpub}, a $\sim 10$ K
hysteresis is observed on the surface. The doping dependence for $T_{c}$ has been evaluated for $0.2 \le x < 0.5$
($x=0.2,0.3,0.4$ and $0.5$) and the general trend is similar to $x = 0.3$: the surface $T_{c}$ is suppressed from
bulk values and a hysteresis is always observed. T* is typically larger than the bulk transition temperature. The
general behavior of this surface phase transition is displayed in Fig.~\ref{PhaseDiagram}.

For $x < 0.2$ a doping-dependent metal-to-insulator transition exists. Upon cooling, a Mott transition occurs
between a paramagnetic metal and an antiferromagnetic insulator. In the bulk, the transition is coupled to a
structural phase transition \cite{nakatsuji_prl00, friedt_prb01, braden_prb98}. While the structural phases for
larger values of $x$ can be described by rotations and tilts of a rigid RuO$_{6}$ octahedron, it is found that in
the insulating state, the octahedron is flattened. The flattening octahedron across the MIT is characterized by a
sharp decrease of the Ru-O(2) bond lengths and increased tilts. On the surface, previous studies have shown the
surface MIT $T_{c}$ to be $\sim 20$ K lower than the corresponding bulk value for $x = 0.1$ and it is imperative
to understand the role of surface structure across the phase boundary \cite{moore_sci_unpub}.
Figs.~\ref{Ca19RuO6}b and ~\ref{Ca19RuO6}c reveal striking deviations between surface and bulk behavior, the
surface structure across the MIT does not change. While the RuO$_{6}$ tilt increases, it does not increase to
those values encountered in the insulating bulk. In addition, the Ru-O(1) basal plane and Ru-O(2) apical bond
lengths, as well as all other structural parameters, remain static through the phase transition. A $3.3${\%}
increase in RuO$_{6}$ volume is encountered due to a $\sim 4^\circ$ increase in RuO$_{6}$ rotation on the surface.
While it has been argued that the structural distortions across the bulk Mott MIT are responsible for the electron
localization \cite{dai_cm06, liebsch_prl07}, the surface MIT is not coupled to any structural phase transition and
is purely electronic in character, i.e. \textit{inherent } \cite{moore_sci_unpub}.

Lower HTT-LTO transition temperatures and the lack of a structural distortion across the surface MIT suggest the
tilt is stabilized on the surface. In addition, LDA calculations reveal the inward motion of the Ca/Sr plane
interferes with the tilting of the RuO$_{6}$ across the $x = 0.1$ MIT \cite{moore_sci_unpub}. The general trend
suggests the inward motion of the topmost Ca/Sr ions plays a significant role in both the static tilt across the
MIT for $x = 0.1$ and the suppressed HTT-LTO phase boundary for $0.2 \le x < 0.5$. The inward motion of the top
Ca/Sr ions creates a compression stress which interferes with the RuO$_{6}$. Theoretical calculations suggest a
similar surface compression should exist on other perovskite material surfaces but experimental evidence has been
lacking \cite{heifets_prb01, piskunov_ss05}. The observed CSRO surface trends would suggest the $x_{c} = 0.5$ bulk
quantum critical point (QCP) should be shifted to lower $x$ on the surface. However, initial results near the QCP
reveal the surface phases to be more complex. While bulk studies reveal the HTT-LTO $T_{c} = 155$ K for $x = 0.4$
\cite{moore_unpub}, a significant surface suppression of the RuO$_{6}$ tilt is encountered as no evidence for the
HTT-LTO transition is observed down to $80$ K. On the contrary, weak diffuse superstructure reflections are
evident at $\sim 80$ K (T*) for $x = 0.5$ on the surface and the HTT-LTO phase boundary is revealed at $T_{c} \sim
40$ K. Extrapolation of both $T_{c}$ and T* to zero in Fig.~\ref{PhaseDiagram} shows that the broken symmetry at
the surface will most likely displace or even destroy the $x_{c} = 0.5$ QCP at the surface. Further investigations
are required to fully determine the existence and position of the QCP on the surface.

In summary, the surface structural phase diagram of Ca$_{2-x}$Sr$_{x}$RuO$_{4}$ has been determined and is
presented in Fig.~\ref{PhaseDiagram}. The RT surface structural phases follow bulk trends with the exception that
no \textit{I4/mmm} symmetry is observed on the surface for $x \ge  1.5$. Significant deviations between surface
and bulk behavior are encountered across $T$-dependent structural phase boundaries. While the RuO$_{6}$ rotation
is revealed for all $x$, a large inward motion of the topmost Ca/Sr ions interferes with the RuO$_{6}$ tilt. As a
result, lower surface HTT-LTO transition temperatures are observed for $0.2 \le x < 0.5$ and the surface Mott MIT
$T_{c}$ is suppressed for $x < 0.2$. In addition, further significant surface deviations from bulk behavior is
noted as a hysteresis is observed across the surface HTT-LTO phase boundary and the structural transitions
accompanying the Mott MIT in the bulk are simply nonexistent on the surface. Implications of the inward motion of
the top Ca/Sr ions on the QCP at $x_{c} \sim  0.5$ are not yet clear as an unexpected HTT-LTO phase boundary is
revealed on the surface at $T_{c} \sim  40$ K for $x = 0.5$.


\begin{acknowledgments}
This work was supported by DOE, Division of Material Science and Engineering through
ORNL. RGM acknowledges support from NSF and DOE (DMS{\&}E) (NSF-DMR-0451163). JZ would like to thank support from
US NSF, under contract DMR-0346826.
\end{acknowledgments}

\end{document}